\documentclass[10pt,letterpaper]{article}
\usepackage[top=0.85in,left=2.75in,footskip=0.75in]{geometry}

\usepackage{changepage}

\usepackage[utf8x]{inputenc}

\usepackage{textcomp,marvosym}

\usepackage{amsmath,amssymb}

\usepackage{cite}

\usepackage{nameref,hyperref}

\usepackage[right]{lineno}

\usepackage{color}

\usepackage{microtype}
\DisableLigatures[f]{encoding = *, family = * }

\raggedright
\usepackage[aboveskip=1pt,labelfont=bf,labelsep=period,justification=raggedright,singlelinecheck=off]{caption}

\makeatletter
\renewcommand{\@biblabel}[1]{\quad#1.}
\makeatother

\date{}

\usepackage{lastpage,fancyhdr,graphicx}
\usepackage{epstopdf}
\fancyheadoffset[L]{2.25in}
\fancyfootoffset[L]{2.25in}

\begin{document}
\vspace*{0.2in}

\begin{flushleft}
{\Large
\textbf\newline{Dynamics of Transformation from Segregation to Mixed Wealth Cities} 
}
\newline
\\
Anand Sahasranaman\textsuperscript{1 2 *},
Henrik Jeldtoft Jensen\textsuperscript{1 2},
\\
\bigskip
\textbf{1} Department of Mathematics and Centre for Complexity Science, Imperial College London, London, United Kingdom
\\
\bigskip

%
%
\textbf{2} Current Address: Centre for Complexity Science, 12th Floor EEE Building, Imperial College London, London, SW7 2AZ, United Kingdom 


\bigskip
* Corresponding Author: Anand Sahasranaman (a.sahasranaman15@imperial.ac.uk)

\end{flushleft}
\section*{Abstract}
We model the dynamics of a variation of the Schelling model for agents described simply by a continuously distributed variable - wealth. Agent movement is not dictated by agent choice as in the classic Schelling model, but by their wealth status. Agents move to neighborhoods where their wealth is not lesser than that of some proportion of their neighbors, the threshold level. As in the case of the classic Schelling model, we find here that wealth-based segregation occurs and persists. However, introducing uncertainty into the decision to move - that is, with some probability, if agents are allowed to move even though the threshold condition is contravened - we find that even for small proportions of such disallowed moves, the dynamics no longer yield segregation but instead sharply transition into a persistent mixed wealth distribution, consistent with empirical findings of Benenson, Hatna, and Or. We investigate the nature of this sharp transformation, and find that it is because of a non-linear relationship between allowed moves (moves where threshold condition is satisfied) and disallowed moves (moves where it is not). For small increases in disallowed moves, there is a rapid corresponding increase in allowed moves (before the rate of increase tapers off and tends to zero), and it is the effect of this non-linearity on the dynamics of the system that causes the rapid transition from a segregated to a mixed wealth state. The contravention of the tolerance condition, sanctioning disallowed moves, could be interpreted as public policy interventions to drive de-segregation. Our finding therefore suggests that it might require limited, but continually implemented, public intervention - just sufficient to enable a small, persistently sustained fraction of disallowed moves so as to trigger the dynamics that drive the transformation from a segregated to mixed equilibrium.

\section*{Introduction}
The Schelling process~\cite{bib1} was developed to explain the emergence of racial segregation in American cities. The central insight of the model was that even a small preference for like neighbors at an individual level can lead to large scale segregation at a collective level. Ever since it was first developed, the basic form of the Schelling model has been widely applied to study the process of segregation using agent-based simulations in sociology~\cite{bib2}~\cite{bib3}~\cite{bib4}~\cite{bib5} and economics~\cite{bib6}. The Schelling model has also increasingly been of interest to physicists, specifically in the application of the tools of statistical mechanics to understand the underlying phase transition occurring in the model~\cite{bib7}~\cite{bib8}~\cite{bib9}~\cite{bib9a}, as also for the development of analytic frameworks of the process~\cite{bib9b}. 

Schelling's original model~\cite{bib1} explored the emergence of segregation for two races, based on the underlying tolerance levels of agents to those of the other race. Race is interpreted as any binary attribute that is unchangeable and easily identifiable. So, for instance, if an agent has a tolerance level of 70\%, it means that the agent would only move into those neighborhoods where at least 30\% of neighbors have the same attribute (or race). The model's dynamics occur in a 'city', which is a one-dimensional or two-dimensional lattice, and an agent's neighborhood in a two-dimensional lattice is typically considered to be either the von Neumann neighborhood of the four nearest neighbors around a location in the lattice or the Moore neighborhood of the eight next nearest neighbors around a location. Schelling finds that even if agents have reasonably high tolerance levels (up to approximately 67\%), large-scale segregation rapidly obtains as agents seek out neighborhoods that satisfy their tolerance levels. Consequently, a highly segregated city ensues despite individual agents being able to tolerate a much more mixed city. The emergence of segregation in the Schelling model has been found to be a robust result across a range of parameter specifications in the model such as lattice size, neighborhood shape and size, and heterogeneity of agent preferences~\cite{bib2}~\cite{bib10}~\cite{bib11}~\cite{bib12}. Segregation is also seen to be robust across a selection of agent choice functions - deterministic, discrete stochastic, or continuous~\cite{bib13}~\cite{bib14}.

In this paper, we explore the dynamics of segregation based on the wealth of agents and are particularly interested in mechanisms that prevent the onset of segregation. There has been some prior work in extending the classical exploration of Schelling's binary-attribute agent model to agents with a wealth or income attribute sampled from a continuous distribution~\cite{bib3}~\cite{bib5}. Benenson, Hatna, and Or~\cite{bib3} attempt to explain observed urban income segregation outcomes in Israel by simulating agent movement dynamics subject to agent incomes (sampled from a log-normal distribution), social utility (tolerance limits), and economic utility (housing affordability), with agents having a view of the entire lattice and choosing to move into the best available site that is available and affordable at any time.  They find not only that income-based segregation occurs and persists as predicted by Schelling, but also that introducing a small number of agents with high tolerance levels into the city could significantly increase the heterogeneity of residential patterns. Benard and Willer~\cite{bib5} model segregation dynamics based on wealth and status of agents (both sampled from a normal distribution) subject to the extent of endogeneity in housing prices and find that wealth based segregation is pronounced with increasing correlation between status and wealth, as also with increasing endogeneity in housing prices. 

While our paper also seeks to explore the dynamics of wealth based segregation, we propose to reorient the classic Schelling model where agent choice typified by agent tolerance levels drives the dynamics of spatial patterns, toward an alternative frame of reference where affordability based on agent wealth determines the dynamics of movement. We believe this reorientation to be of considerable interest because it could be reasonably argued that affordability, even more than desire, drives housing choice in real world scenarios.  Essentially, this means shifting the perspective of agent dynamics away from the endogenous tolerances of individual agents to an exogenous and universal affordability threshold applied to all agents. As in the classic Schelling case, we expect to see the emergence of segregation, and hope that our contribution will be in the exploration of mechanisms to reverse or prevent the onset of segregation.

Some of the recent physics literature~\cite{bib9}~\cite{bib9a} explores the interface between segregation and de-segregation and suggests mechanisms to enable de-segregation. Hazan and Randon-Furling~\cite{bib9} modify Schelling dynamics for two types of agents by introducing a certain fraction of 'switching' agents that can change from one type to another over the course of time and find that the presence of switching agents significantly reduces the extent of segregation. Grauwin, Bertin, Lemoy, and Jensen~\cite{bib9a} introduce the notion of collective utility in addition to individual utility, basing their dynamics on a combination of collective and individual utility (calibrated by a factor called the 'degree of cooperativity') and find a critical value of 'degree of cooperativity' between mixed and segregated states. Our paper furthers the spirit of these studies and explores fundamental mechanisms to enable mixed wealth equilibria.

\section*{Model Definition and Specifications}
We seek to understand the evolution of spatial dynamics on account of the wealth of individual agents in a city. Each individual agent can be thought of as a household that resides in a single cell on a two-dimensional lattice with $N$ cells that represents the city. The neighborhood of a cell is defined to be its eight next nearest (or Moore) neighbors. 

Each agent has a single attribute, namely its wealth, which forms the basis for the dynamics we explore. The actual spreads of incomes and wealth in nations around the world are described by a range of distributions - exponential, log-normal, gamma, and power-law~\cite{bib17}~\cite{bib18}~\cite{bib19}~\cite{bib20}. The wealth of agents is sampled from a standard normal distribution for the purpose of this paper, but our findings are also robust for exponential and log-normal distributions. These agent wealths are initially distributed randomly across locations in the lattice. 

As previously discussed, we re-interpret the tolerance level in the original Schelling model to a 'threshold level', $\tau$, defined as the minimum number of neighbors of a site whose wealth has to be lesser than or equal to the wealth of any agent attempting to move in to that site. Instead of an agent's tolerance in the case of the classic Schelling model, in this case we conceptualize the threshold level to be defined at the level of the lattice (city) such that it is the uniform requirement for every agent to move locations within the city. Low $\tau$ values indicates a higher potential for movement for all agents across the wealth spectrum, while a high $\tau$ indicates much lesser potential for movement for those at the lower end of the wealth spectrum. The level of $\tau$ can therefore be considered an apriori measure of the ease of movement in the city based on an agent's wealth. For the model, we assume that an agent can move to a new location only if her wealth is greater than or equal to the wealth of half of the eight Moore neighbors in the new location. Therefore, we set $\tau$ = 4. While this is the value of $\tau$ for the simulation of agent dynamics, we are also interested in varying the value of $\tau$ to understand if there is a threshold for segregation to emerge. In the case of the classic Schelling two-race model, segregation obtains up to a tolerance level $\simeq2/3$. Above this tolerance level there is no segregation as the high tolerance of all agents ensures a mixed spatial distribution. In our model as well, we find that there indeed does exist a threshold: for $\tau\le4$, segregation emerges and persists. This corresponds to a threshold ratio (defined as $\tau$/8) of 50\% of neighbors for the emergence of segregation. Above this threshold ratio, segregation does not emerge. 
 
The dynamics of the model are generated through Kawasaki kinetics~\cite{bib21}, where two sites, with agents A and B, are picked at random from the lattice. Each selected agent's wealth is compared to the wealth of each of the agents in the other's neighborhood. Let n\textsubscript{A} and n\textsubscript{B} be the number of neighbors of A(B) whose wealths are lesser than or equal to that of B(A) respectively. If both n\textsubscript{A} and n\textsubscript{B} are greater than or equal to $\tau$, then the agents exchange places with probability 1. However, if this is not the case, then an exchange occurs with probability p\textsubscript{m}. (Eq~\ref{eq:probmove})
\begin{eqnarray}
\label{eq:probmove}
	\mathrm{p_m} = {\exp (\beta\Delta)} \\
     where: \mathrm{\Delta} = {n_A + n_B - 4\tau}
\end{eqnarray}

While we have chosen an exponential form for the probability of move (p\textsubscript{m}) function, the dynamics appear robust to choice of functional form. For instance, we find that the results of the simulation can be replicated using the power law functional form as well.

In each iteration of the model, two agents are picked at random and depending on whether they satisfy the threshold condition, the agents exchange location based on the mechanism described earlier. We run 100,000 iterations for each realization of the dynamics with a given value of $\beta$, and repeat this 100 times for each $\beta$. All the measures to calculate the extent of segregation (described next) are averaged over these 100 realizations. This process is repeated for the entire range of $\beta$ values. We choose 100,000 iterations because we find that for the chosen lattice size of $N$ = 625, the statistical properties of the spatial configuration capturing the extent of segregation settle down and remain stable beyond this number.

We use two measures to capture the extent of segregation obtained in the lattice as a consequence of the dynamics: Size of Rich Neighborhoods (S) and Rich Wealth Differential ($\delta$). In this context, 'rich' is defined as the set of agents with wealth greater than or equal to one positive standard deviation from the mean of the wealth distribution.  Let N\textsubscript{r} be the number of rich agents in the wealth distribution. For the standard normal distribution, the size of this set is 15-16\% of the total agent population.

Size of Rich Neighborhoods (S) is simply a measure of the average number of rich neighbors of a rich agent. For each rich cell $i$, we compute the number of its neighbors that are also rich, V\textsubscript{r}(i) (Eq~\ref{eq:Si}). S is the average of this quantity across all rich cells (~\ref{eq:S}). We expect S to increase with increasing segregation on account of the spatial congregation of the rich in more segregated states.
\begin{eqnarray}
\label{eq:Si}
	\mathrm{S_i} =  {\Sigma_{j\in V_r(i)}   1} 
\end{eqnarray}
\begin{eqnarray}
\label{eq:S}
	\mathrm{S} = \frac{\Sigma_{i} S_i}{N_r}
\end{eqnarray}

Rich Wealth Differential ($\delta$) is a measure of the average absolute difference in wealth between a rich cell and its neighbors. For each rich cell $i$, we calculate $\delta$\textsubscript{i}, which is the average of the absolute difference between the wealth of $i$ and the wealth of each of its Moore neighbors (Eq~\ref{eq:deli}). $\delta$ is obtained by averaging $\delta$\textsubscript{i} for all rich cells (Eq~\ref{eq:del}). As segregation increases, we would expect $\delta$ to decrease. This is because, as neighborhoods become more homogeneous with increasing segregation, we expect that, on average, the absolute difference in wealth between rich cells and their neighbors will progressively decrease.
\begin{eqnarray}
\label{eq:deli}
	\mathrm{\delta_i} = \frac{\Sigma_{j\in V(i)} \vert W_i - W_j \vert}{8}
\end{eqnarray}
\begin{eqnarray}
\label{eq:del}
	\mathrm{\delta} = \frac{\Sigma_{i} \delta_i}{N_r}
\end{eqnarray}
In Eq~\ref{eq:deli}, V(i) is the Moore neighborhood of cell $i$. 

Table~\ref{table1} summarizes the basic model parameters: \newline

\begin{table}[!ht]
\centering
\captionsetup{justification=centering}
\caption{\bf Model Parameters.}
\begin{tabular}{|l|l|}
\hline
Lattice Size & 25 X 25\\ \hline
Number of Cells ($N$) & 625\\ \hline
Wealth Distribution & N(0,1)\\ \hline
Threshold Level ($\tau$) & 4 \\ \hline
Neighborhood & Moore \\ \hline
$\beta$ & 0.001 - 100 \\ \hline
\end{tabular}
\label{table1}
\end{table}

In order to ensure that our simulation results are not predicated on the specific geometry chosen - a two-dimensional square lattice with a Moore neighborhood - we also alternatively model the spatial dynamics on a random graph using the Erd\H{o}s-R\'enyi model~\cite{bib21a}. Nodes in the graph represent agents and edges of a node define the node's neighborhood. We find that the results from the random graph model are similar to the results of the simulations from a lattice model. 

\subsection*{Comparison with the Classic Schelling Model}
Finally, we simulate the classic Schelling model of wealth based segregation to compare the results with our variant of the model. Modeling the classic Schelling version where the choice of move is determined by the desire of an agent to be in a neighborhood where at most T agents are of lesser wealth (T being the tolerance level of agents), we find that that as $\beta\rightarrow0$, this model also exhibits similar behavior to that simulated by our variant of the model. We also explore the emergence of segregation with changing T. 

While the nature of results produced by these alternate models is similar, the two models do however provide differing perspectives on understanding segregation. The classic Schelling version is one where agent choice (tolerance) drives the spatial configurations and where a decreasing $\beta$ could be interpreted as a secular increase in the tolerance of all agents or the presence of some fraction of agents with high levels of tolerance. Our variant of the Schelling model can be interpreted as the dynamics resulting from thresholds that cities present to agents (wealth thresholds), where a decreasing $\beta$ can be understood as public policy measures that enable greater ease of movement and drive heterogeneous spatial configurations.

\section*{Description of Emergent Dynamics}
For $\beta\ge5$, irrespective of the value of $\Delta$, p\textsubscript{m} is so small that no disallowed-realized moves occur and the only moves that occur are those that satisfy the threshold condition. At the other end of the spectrum, for $\beta\simeq0.001$, disallowed-realized moves form 68\% of all moves that occur.

\begin{figure}[!h]
\centering
\includegraphics[width=0.75\textwidth]{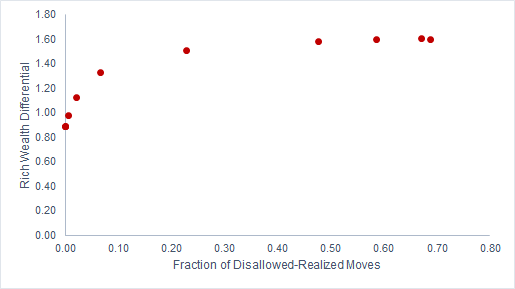}
\caption{{\bf Rich Wealth Differential v. Fraction of Disallowed-Realized Moves.}}
\label{fig1}
\end{figure}

Simulating the dynamics we find a non-linear relationship between Rich Wealth Differential ($\delta$) and disallowed-realized moves. At very high $\beta$, when no disallowed-realized moves occur, $\delta$ is at a minimum, but as $\beta$ is decreased and disallowed-realized moves just begin to occur, we find that there is a sharp increase in $\delta$ and this continues till the fraction of disallowed-realized moves reaches about 10\% (or a $\beta$ in the range 0.25 to 0.5). Beyond this region, there is a more gradual increase in $\delta$ till it stabilizes for $\beta\le0.05$, corresponding to disallowed-realized moves ratios in the region of 58-68\%. The overall increase in $\delta$ over the entire range of $\beta$ is just over 80\% (Fig~\ref{fig1}). However, it is in the initial region, where the fraction of disallowed-realized moves is less than 10\%, that most of this increase occurs. In fact, over 62\% of the increase in $\delta$ occurs even as the fraction of disallowed-realized moves is under 6.5\%. 

\begin{figure}[!h]
\centering
\includegraphics[width=0.75\textwidth]{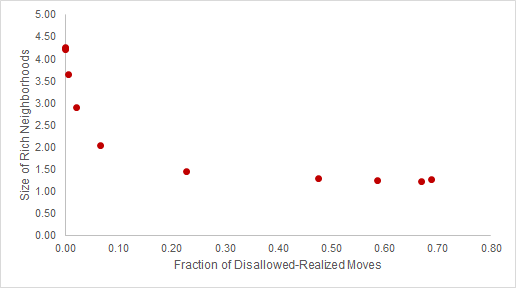}
\caption{{\bf Size of Rich Neighborhoods v. Fraction of Disallowed-Realized Moves.}
}
\label{fig2}
\end{figure}

A similar non-linear relationship is observed in the relationship between Size of Rich Neighborhoods (S) and the fraction of disallowed-realized moves(Fig~\ref{fig2}). For $\beta\ge5$, where no disallowed moves occur, S is in the region of 4.25, meaning that the top 15\% of the wealthiest agents have, on average, 4.25 like neighbors. However, there is again a sharp transition downwards as disallowed-realized moves begin to occur and S drops to 2 as the fraction of disallowed-realized moves increases to a mere 6.5\% - a drop of over 74\% (of the overall drop), which is similar to the sharp increase we saw in the case of $\delta$ earlier. Again, beyond this point, the curve exhibits a gentler decline and size of rich neighborhoods settles in the region of 1.25 for $\beta\le0.05$, resulting in an overall drop of 70\% in S. 

\begin{figure}[!h]
\centering
\includegraphics[width=0.75\textwidth]{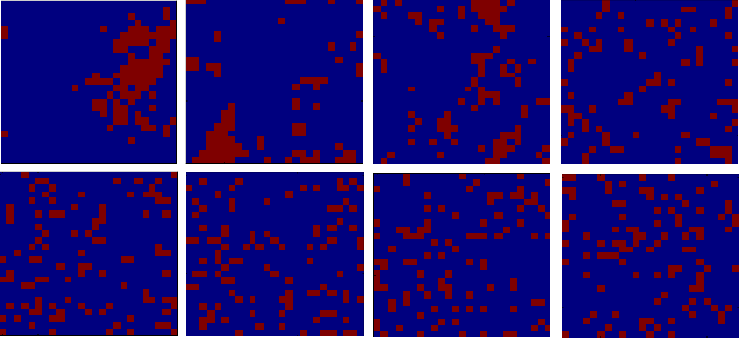}
\caption{{\bf Sample Heat-map of the Extent of Rich Neighborhoods.}
Legend: D-R moves = Disallowed-Realized moves, S = Size of Rich Neighborhoods. A: $\beta$ = 5, D-R moves = 0\%, S = 4.5 (top left). B: $\beta$ = 1, D-R moves = 0.46\%, S = 3.6 (top 2\textsuperscript{nd} left). C: $\beta$ = 0.75, D-R moves = 2.04\%, S = 2.8 (top 2\textsuperscript{nd} right). D: $\beta$ = 0.5, D-R moves = 6.91\%, S = 1.6 (top right). E: $\beta$ = 0.25, D-R moves = 22.62\%, S = 1.4 (bottom left). F: $\beta$ = 0.1, D-R moves = 47.50\%, S = 1.2 (bottom 2\textsuperscript{nd} left). G: $\beta$ = 0.05, D-R moves = 58.18\%, S = 1.3 (bottom 2\textsuperscript{nd} right). H: $\beta$ = 0.001, D-R moves = 68.81\%, S = 1.4 (bottom right).}
\label{fig3}
\end{figure}

Fig~\ref{fig3} captures the transformation from a segregated to mixed wealth state using a heat-map. These heat-maps represent one particular realization of dynamics post 100,000 iterations, across a range of $\beta$ values. As is apparent, at very high $\beta$, the rich cells (in red) are surrounded by many rich cells - the large cluster of red for $\beta$ = 5 corresponds to S = 4.5. However, as soon as disallowed-realized moves start happening, we observe the size of rich neighborhoods rapidly decreasing. At  $\beta$ = 1, when the fraction of disallowed-realized moves is merely 0.46\%, S has dropped to 3.6.  When the disallowed-realized moves ratio increases to 6.9\% ($\beta$ = 0.5), S drops substantially further to 1.6. As the disallowed-realized moves ratio increases further, there is a much gentler drop off in S as it settles to a value in the region of 1.3 for $\beta\le0.05$. 

\begin{figure}[!h]
\centering
\includegraphics[width=0.75\textwidth]{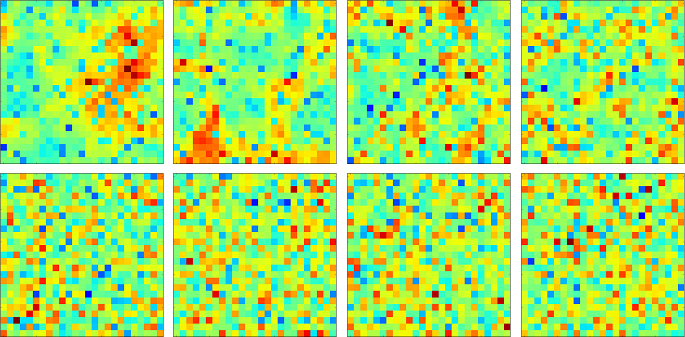}
\caption{{\bf Sample Continuous Value Heat-maps of the Extent of Neighborhoods.}
Legend:A: $\beta$ = 5(top left). B: $\beta$ = 1 (top 2\textsuperscript{nd} left). C: $\beta$ = 0.75 (top 2\textsuperscript{nd} right). D: $\beta$ = 0.5 (top right). E: $\beta$ = 0.25 (bottom left). F: $\beta$ = 0.1 (bottom 2\textsuperscript{nd} left). G: $\beta$ = 0.05 (bottom 2\textsuperscript{nd} right). H: $\beta$ = 0.001 (bottom right).}
\label{fig3b}
\end{figure}

The complete continuous value heat-maps displaying all agent wealths without isolating just the rich agents is presented in Fig~\ref{fig3b} and this shows the separate spatial congregations, at $\beta$ = 5, of both the richest (in orange clusters) and poorest (in bluish clusters) agents, yielding segregation. Lower $\beta$ values indicate more mixed configurations and progressive de-segregation.

Clearly, there is a sharp transformation from a segregated state where there is no potential for disallowed moves to a substantially more mixed state even as there is only a small increase in fraction of disallowed-realized moves. This sharp transformation is fully consistent with the findings of Benenson, Hatna, and Or~\cite{bib3} who observe heterogeneity in the spatial distribution of wealth in some Israeli cities and explain it as a consequence of the high tolerance levels of a small proportion of rich households. For instance, their modeling of observed residential patterns leads them to conclude that introducing only 2.5\% of agents with high tolerance levels leads to an essential increase in heterogeneity, irrespective of how high the levels of intolerance of the remaining 97.5\% of agents are. At 10-15\% tolerant agents, they find that heterogeneity is as high as it can possibly be.

We now explore the nature of this transformation from a segregated to a mixed state, and propose a possible mechanism that drives this phenomenon.

\section*{Discussion}
\subsection*{Sharp Transformation from Segregated to Mixed State}
\begin{figure}[!h]
\centering
\includegraphics[width=0.75\textwidth]{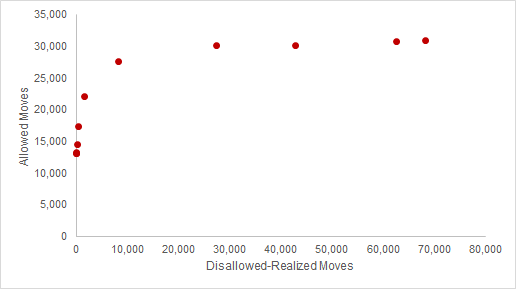}
\caption{{\bf Number of Allowed Moves v. Number of Disallowed-Realized Moves.}
}
\label{fig4}
\end{figure}

In order to understand the transformation from a segregated to a mixed state, we start by exploring the variation in allowed moves as $\beta$ is varied. Allowed moves are moves that occur when the threshold condition is satisfied for both agents in an iteration. When there are no disallowed moves ($\beta\ge5$), the number of allowed moves is around 13,000. As soon as disallowed moves start being realized, the total number of moves increase not just by the number of disallowed-realized moves, but by additional allowed moves that are generated on account of these disallowed-realized moves. As Fig~\ref{fig4} illustrates, just as disallowed-realized moves start occurring (for $\beta\le1$), there is a dramatic increase in allowed moves. When the number disallowed-realized shows a small increase from 0 to 80, the corresponding increase in allowed moves is in the region of 1,500. While this dramatic effect on allowed moves is pronounced for the early increases in disallowed-realized moves (up to about 1,500 moves), beyond this initial region there is a rapid decrease in the rate of increase in allowed moves until finally the number of allowed moves settles (and remains invariant to further changes in disallowed-realized moves) at a value of 30,000-31,000.

The non-linear explosion in the number of allowed moves as function of the number of disallowed-realized moves involves  what may be called an entropic effect. This is caused by a very rapid increase in the number of ways to select pairs that satisfy the threshold condition. To see how this works imagine the lattice is fully occupied by a population of two types. Members of one community all have the same wealth $W_2>W_1$ and the members of the other all have wealth $W_1$. Now consider a region A, where a square cluster of $W_2$ agents is surrounded by $W_1$ agents. Except for the four cells at the vertices of the square (which for large clusters will be selected with vanishingly small probabilities), the $W_2$ agents at the edge of the $W_2$ cluster cannot be swapped with $W_1$ agents through allowed moves. But now assume that one $W_1$ agent has been swapped through a disallowed-realized move to one of the edge sites within the $W_2$ cluster. Two neighbor sites of this site now each have 4 neighbors with wealth $W_2$ and four with wealth $W_1$. Thus these two sites have become available for allowed swaps for the $W_1$ agents.  So one disallowed-realized move has generated two new sites for allowed moves.    
 
Phrased in more general terms, even as a very small proportion of disallowed moves begins to happen, the resulting configuration of agents on the lattice enables more allowed moves to happen. The effect of a decreasing $\beta$ is that neighborhoods that previously would not have had lower wealth agents moving in on account of the threshold condition now see such disallowed moves, even if in small numbers initially. Consequently, other agents on the lattice whose wealth is in the lower end of the distribution and who were earlier unable to move because they were close, but unable, to achieving the threshold condition are now able to achieve it on account of the low wealth agents that have moved around in the lattice to the neighborhoods that were out of reach. This effect is especially pronounced at lower fractions of disallowed-realized moves, where even small increases in such moves yield substantial increases in allowed moves as many agents that were earlier close to fulfilling the threshold levels are able to move now. 

It is this cascade in allowed moves caused by small numbers of disallowed moves that yields the sharply declining segregation we observed earlier. However, as $\beta$ is decreased, we find that the increasing fraction of disallowed-realized moves has a progressively weaker effect on allowed moves, until a point is reached where allowed moves settle at a stable value and stop varying with disallowed-realized moves. This tapering off effect is again consistent with the observed gradual decline in segregation after the initial sharp decline. This non-linear behavior of allowed moves could be explained by the fact that once a certain level of disallowed moves are realized, the lattice reaches mixed wealth configurations that maximize allowed moves, and any further increase in disallowed-realized moves has no more impact on allowed moves.

\begin{figure}[!h]
\centering
\includegraphics[width=0.75\textwidth]{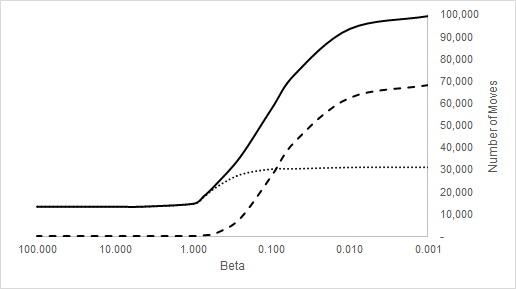}
\caption{{\bf Change in Absolute Number of Moves with $\beta$.}
Legend: Solid Line = Total Realized Moves, Dotted Line = Allowed Moves, Dashed Line = Disallowed-Realized Moves.}
\label{fig5}
\end{figure}

Fig~\ref{fig5}, which is a plot of the absolute number of moves (total, allowed, and disallowed-realized) confirms that at high $\beta$, we observe no disallowed moves, but just when $\beta\simeq1$ and the first disallowed-realized moves manifest, the number of allowed moves starts to increase as well. This increase is rapid in the region between $\beta$ = 1 and $\beta$ = 0.1. At $\beta\simeq0.1$, the number of disallowed-realized moves crosses the number of allowed moves and it is at this point that the value of allowed moves stabilizes. All increases in total moves beyond this point are on account of change in disallowed-realized moves alone.

\begin{figure}[!h]
\centering
\includegraphics[width=0.75\textwidth]{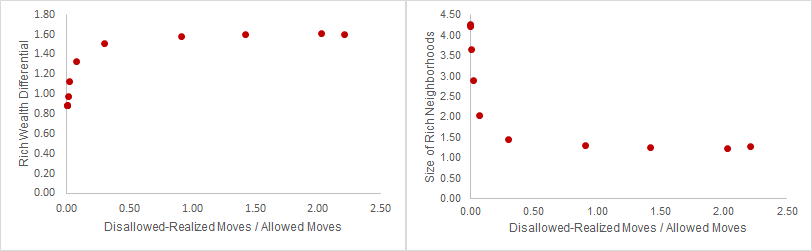}
\caption{{\bf Neighborhood Statistics.}
A: Rich Wealth Differential v. Disallowed-Realized Moves / Allowed Moves (left). B: Size of Rich Neighborhoods v. Disallowed-Realized Moves / Allowed Moves (right).}
\label{fig6}
\end{figure}

If we now consider changes in $\delta$ and S as a function of the ratio of disallowed-realized moves to allowed moves, we would expect to see the sharp transformation from segregated to mixed states occurring when this ratio is small and then level off as the ratio reaches 1. Fig~\ref{fig6} confirms this behavior.

\subsection*{Potential Public Policy Implications}
There is a significant body of literature in economics that analyzes the effects of economic segregation in cities on a wide range of social outcomes~\cite{bib22}~\cite{bib23}~\cite{bib24}. Wealth or income based segregation has been shown to adversely impact the educational attainment of youth from poorer households~\cite{bib22}. Levels of civic participation tend to be significantly lower in economically homogeneous cities than in economically diverse, middle income cities~\cite{bib23}. Additionally, there is evidence that socioeconomic segregation significantly impacts the risk of mortality, especially among the poor~\cite{bib24}. In view of these adverse outcomes of segregation, the prevention or reversal of segregation in cities remains a question of significant import to public policy. 

While the model of segregation presented in this paper is a stylistic one, our results do indicate that the objective of long term de-segregation cannot be achieved by significant one-time public investments, but rather through minimal, dynamically implemented strategies that continually enable small fractions of households to move into areas that they may otherwise be unable to afford. The significant public policy challenge is therefore to devise strategies that allow for such continuous mixing over time rather than to focus on static strategies that might enable mixing over short time frames (temporarily lowering $\beta$), before their effect subsides and segregation obtains again.The dynamic application of policy interventions appears essential to effectively respond to emerging patterns of settlement and consequently enable long term persistence of mixed equilibria.

\section*{Conclusion}
We apply a variant of the Schelling model to the wealth of agents in a city. As expected, we find that subject to a threshold level, wealth based spatial segregation ensues. However, introducing uncertainty into the decision to move, and randomly allowing some proportion of agents to move even in contravention of the threshold level condition, we find that even for small proportions of such disallowed moves, the dynamics no longer settle into a segregated state but instead sharply transition into a mixed wealth distribution. We find that this sharp transformation is on account of a non-linear increase in the number of allowed moves that occurs as a consequence of the disallowed-realized moves. 

The contravention of the tolerance condition could be interpreted as minimal, but dynamically implemented, public policy interventions to enable the continuous realization of a small fraction of disallowed movement. The dynamics generated by such moves could aid the transformation from segregated to mixed cities.

\nolinenumbers

\end{document}